\documentclass[prl,aps,showpacs,twocolumn]{revtex4}

\usepackage{graphicx}

\begin{document}

\title{Fractal structure of the effective action in (quasi-) planar
models with long-range interactions}
\author{E.V. Gorbar$^{1*}$, V.P. Gusynin$^{2*}$, V.A.~Miransky$^{2*}$,
and I.A. Shovkovy$^{3*}$  }
\affiliation{$^{1}$ Universidade Federal de Juiz de Fora,
Juiz de Fora 36036-330, Brazil\\
$^{2}$ Department of Applied Mathematics, University of Western
Ontario, London, Ontario N6A 5B7, Canada\\
$^{3}$ Institut f\"{u}r Theoretische Physik, 
       J.W. Goethe-Universit\"{a}t, 
       D-60054 Frankurt/Main, Germany 
}


\begin{abstract}
We derive the effective potential for composite fields in
a class of (quasi-) planar models with long-range interactions. 
This class of models can be relevant for  
high temperature superconductors and graphite. The fractal structure 
of the effective potential is revealed and its physical interpretation
is presented. It is argued that the multi-branched fractal structure 
of the potential reflects the presence of an infinite tower of 
excitonic bound states that occur as a result of the long-range
interactions.
\end{abstract}


\maketitle

For a long time, there has been a great interest in low dimensional
quantum field theories. Typical examples are the $1+1$ dimensional
Gross-Neveu model and the $2+1$ dimensional quantum electrodynamics
(QED$_{2+1}$).  They have been used as very instructive toy models of
relativistic systems with complicated dynamics. In addition to that, it is
quite remarkable that these same models mimic some intrinsic properties of
the dynamics of non-Fermi liquids. In particular, numerous recent studies
suggest that high-$T_c$ superconductors and graphite could be described
well by a (quasi-) planar QED$_{2+1}$
\cite{highTc,kim,highTc2,Semenoff,carbon,Khvesh,HOPG}.

The difficulties in studying strongly correlated systems such as
high-$T_c$ superconductors and graphite might be related in part to the
presence of long range Coulomb like interactions between
quasiparticles. This limits considerably the use of many
powerful techniques that used to work well in models with short-range
interactions.  One of such tools is the celebrated Ginzburg-Landau (GL)
effective action. A method of the derivation of the GL action for
composite operators in systems with short-range interactions has been
known since the work of Gorkov \cite{Gork}. Its generalization to
systems with a spontaneous symemtry breaking driven by long-range
interactions is far from being straightforward. In relativistic field
theories, a method of the derivation of the effective action in models
with long-range interactions (driven by gauge fields) was elaborated in
Ref.~\cite{Mir}.

Recently, using the method of Ref.~\cite{Mir}, the effective potential has
been derived in the so-called reduced QED \cite{HOPG}. In such a model,
while quasiparticles are confined to a 2-dimensional plane, the
electromagnetic interaction between them is three dimensional in nature.
The model is relevant for a class of condensed matter systems that include
highly oriented pyrolytic graphite among others \cite{Khvesh,HOPG}. In
this letter, we will reveal a remarkable property of the effective
potential in reduced QED, not considered in Ref.~\cite{HOPG}. It will be
shown that the potential has a multi-branched fractal structure reflecting
the presence of an infinite tower of excitonic bound states (a clear
signature of long-range interactions). It is also important that, as was
shown in Ref.~\cite{HOPG}, reduced QED is intimately connected with
QED$_{2+1}$, relevant for cuprates \cite{highTc,kim,highTc2}. In fact, as
will be discussed below, the qualitative features of the effective
potential in reduced QED are common for a wide class of models with
long-range interactions, in particular, for QED$_{2+1}$.

We start from a general definition of the effective action in a theory in
which the spontaneous symmetry breaking phenomenon is driven by the local
composite order parameter $\langle0|\bar{\psi}\psi|0\rangle$, where $\psi$
is the Dirac spinor of the quasiparticle field and $\bar{\psi}$ is the
(Dirac) conjugate spinor. Following the conventional way (see for example
Ref.~\cite{book}), we introduce the generating functional $W(J)$ for the
Green functions of the corresponding composite field through the path
integral:
\begin{eqnarray}
e^{iW(J)}
&=&\int D\psi D \bar{\psi} 
\exp\Bigg\{i
\int d^3x \left[{L}_{qp}(x)
\right.\nonumber\\  
&-&\left.J(x) \bar{\psi}(x)\psi(x)\right]\Bigg\},
\label{genfunc}
\end{eqnarray}
where $J(x)$ is the source for composite field and ${L}_{qp}(x)$ is
the lagrangian density of quasiparticles in the model at hand.  Then, by
definition, the effective action for the field $\sigma(x)  \equiv
-\langle0|\bar{\psi}\psi(x)|0\rangle$ is given by the Legendre 
transform of the generating functional $W(J)$,
\begin{equation}
\Gamma(\sigma)=W(J)-\int d^3x J(x)\sigma(x),
\label{Gamma}
\end{equation}
where the external source $J(x)$ on the right hand side is expressed
in terms of the field $\sigma(x)$ by inverting the relation
\begin{equation}
\frac{\delta W}{\delta J(x)}=\sigma(x).
\label{dW/dJ}
\end{equation}
The effective action $\Gamma(\sigma)$ in Eq.~(\ref{Gamma}) provides 
a natural framework for describing the low energy dynamics in the 
model at hand. It is common to expand this action in powers of 
space-time derivatives of the field $\sigma$:
\begin{equation}
\Gamma(\sigma)=\int d^3x\left[-V(\sigma)+
\frac{1}{2}Z^{\mu\nu}(\sigma) \partial_\mu
\sigma\partial_\nu\sigma +\dots\right],
\end{equation}
where $V(\sigma)$ is the efective potential. The ellipsis denote higher
derivative terms as well as contributions of the Nambu-Goldstone
bosons if the latter are required by the Goldstone theorem.  By making use
the definition in Eqs.~(\ref{Gamma})  and (\ref{dW/dJ}), we derive the
following relation:
\begin{equation}
\frac{\delta \Gamma}{\delta\sigma(x)}=-J(x).
\label{source}
\end{equation}
In the limit of a vanishing external source, this equation turns into 
an equation of motion for the composite field $\sigma(x)$. 
In a particular case of constant configurations, the equation 
reads $dV/d\sigma =0$. 

By keeping the external source on the right hand side of
Eq.~(\ref{source}) nonzero but constant in space-time, we derive the
following convenient representation for the effective potential:
\begin{equation}
V(\sigma) = -w(J)+ J \sigma = \int^{\sigma} d\sigma J(\sigma),
\label{eff-pot-def}
\end{equation}
where $w(J)\equiv W(J)/V_{2+1}$, and $V_{2+1}$ is the space-time volume.
Thus, the crucial point for evaluating $V$ is how the source $J$, playing
here the role of the ${\it bare}$ fermion gap, depends on the composite
field $\sigma$. It will be shown below, following the approach of 
Ref.~\cite{Mir}, that the function $J(\sigma)$ can be determined from 
a modified gap equation, including the bare gap $J$.

The model under consideration is the same as the model used in
Ref.~\cite{Khvesh,HOPG} to study a metal-insulator phase transition in
highly
oriented pyrolytic graphite. We assume that the fermionic quasiparticles
are confined to a flat $2$-dimensional plane, while the electromagnetic
field is free to propagate in the $3$-dimensional bulk. The spatial
coordinates on a plane are denoted by $\vec{r}=(x,y)$ and the orthogonal
direction is labeled by the $z$ coordinate.

The lagrangian density of the electromagnetic field (in the bulk) 
is given by
\begin{eqnarray}
{L}_{em} &=& \frac{1}{8\pi} \left(\varepsilon_{0} \vec{E}^{2}
-\frac{1}{\mu_{0}} \vec{B}^{2} \right) 
-A_{0} \rho +\frac{1}{c} \vec{A}\cdot \vec{j},
\label{L-1}
\end{eqnarray}
where $\varepsilon_{0}$ is the dielectric constant, 
$\mu_{0}$ is the magnetic permeability,  
$A_{0}$ and $\vec{A}$ are the scalar and vector potentials, and
electric and magnetic fields are 
$\vec{E} = - \vec{\nabla} A_{0} -\frac{1}{c} \partial_{t} \vec{A}$,
$\vec{B} = \vec{\nabla} \times \vec{A}$.
We note that the interaction terms, with the quasiparticle charge 
density $\rho$ and current $\vec{j}$, were explicitly included in 
the lagrangian density in Eq.~(\ref{L-1}). In addition to that, the 
lagrangian density of quasiparticles themselves (living only on the 
plane) should be specified. It reads 
\begin{equation}
{L}_{0} = v_{F}
\bar{\psi}(t,\vec{r})
\left({i\gamma^{0}\partial_{t}}/{v_{F}} 
-i\gamma^{1} \partial_{x}
-i\gamma^{2} \partial_{y}\right)\psi(t,\vec{r}),
\label{L-free}
\end{equation}
where $v_{F}$ is the Fermi velocity,
$\psi(t,\vec{r})$ is a 4-component spinor, $\bar{\psi}=
\psi^{\dagger}\gamma^{0}$, and the $4\times 4$ Dirac $\gamma$-matrices
furnish a reducible representation of the Clifford (Dirac) algebra in
$2+1$ dimensions. 

We consider the case when the fermion fields carry an additional,
``flavor", index $i=1,2,\dots,N_f$. Then, the symmetry of the lagrangian
(\ref{L-free}) is $U(2N_{f})$. Adding a bare quasiparticle gap (mass) term
$\Delta_{\rm br}\bar\psi\psi$ into the lagrangian density (\ref{L-free}) 
would reduce the
$U(2N_{f})$ symmetry down to the $U(N_f)\times U(N_f)$. Therefore the
dynamical generation of the gap would lead to the spontaneous breakdown 
of the $U(2N_f)$ down to the $U(N_f)\times U(N_f)$.

Proceeding as in Ref.~\cite{HOPG}, we integrate out the bulk 
gauge bosons from the action. Then, up to relativistic corrections 
of order $(v_{F}/c)^{2}$, we are left with the following action of 
interacting planar quasiparticles:
\begin{eqnarray}
S_{qp} &\simeq& \int dt d^{2} \vec{r} {L}_{0}(t,\vec{r})
-\frac{1}{2} 
\int  dt d^{2} \vec{r}  
\int dt^{\prime} d^{2} \vec{r}^{\prime}\bar{\psi}(t,\vec{r})
 \nonumber \\
&\times& \hspace{-2mm}\gamma^{0}\psi(t,\vec{r}) 
U_0(t-t^{\prime},|\vec{r}-\vec{r}^{\prime}|)
 \bar{\psi}(t^{\prime},\vec{r}^{\prime})\gamma^{0}
\psi(t^{\prime},\vec{r}^{\prime}),
\label{action}
\end{eqnarray}
where the bare potential $U_{0}(t,|\vec{r}|)$ takes the following 
simple form:
\begin{equation}
U_{0}(t,|\vec{r}|) = \frac{e^2\delta(t)}{\varepsilon_{0}} 
\int \frac{d^{2}\vec{k}}{(2\pi)^{2}}  
\exp(i\vec{k}\cdot\vec{r})\frac{2\pi}{|\vec{k}|}
=\frac{e^{2}\delta(t)}{\varepsilon_{0}|\vec{r}|}.
\end{equation}
Note, however, that the polarization effects may considerably 
modify this bare Coulomb potential. Thus, after taking this into 
account, the interaction is given by
\begin{equation}
U(t,|\vec{r}|) = \frac{e^2}{\varepsilon_{0}}
\int\frac{d\omega}{2\pi} \int \frac{d^{2}\vec{k}}{2\pi}
\frac{\exp(-i\omega t+i\vec{k}\cdot\vec{r})}
{|\vec{k}|+\Pi(\omega,|\vec{k}|)},
\label{retard}
\end{equation}
where the polarization function $\Pi(\omega,|\vec{k}|)$ is 
proportional (with a factor of $2\pi/\varepsilon_{0}$) to
the time component of the photon polarization tensor.

The starting point in our derivation of the effective potential for 
the composite field $\sigma$ is the gap equation in a model 
with a nonzero constant external source $J$ [see Eq.~(\ref{genfunc})]. 
By noticing that the external source term $J\bar{\psi}\psi$ enters 
the action in exactly the same way as the quasiparticle bare mass 
term, we easily derive the corresponding gap equation \cite{HOPG},
\begin{eqnarray}
\Delta_p - J &=&
\lambda\int\frac{qdq\Delta_q{K}(p,q)}{\sqrt{q^2+(\Delta_q/v_F)^2}},
\label{gap_eq:nonlin} \\
\lambda &=& \frac{e^{2}}{2(\varepsilon_{0}v_{F}+\pi e^{2}N_{f}/4)},
\nonumber
\end{eqnarray}
where the approximate expression for the kernel ${K}(p,q)$ is given
by
$
{K}(p,q)={\theta(p-q)}/{p}+{\theta(q-p)}/{q}.
$
In the most important region of momenta $|{\vec q}|\gg \Delta_q/v_F$ where
the pairing dynamics dominates, the only role of the term
$(\Delta_q/v_F)^2$ in the denominator of the integrand on the right-hand
side of Eq.(\ref{gap_eq:nonlin}) is to provide a cutoff in the infrared
region. Therefore, one can drop this term, instead introducing the
explicit infrared cutoff $\Delta/v_F$ in the integral, where the gap
$\Delta$ is defined from the condition $\Delta \equiv
\Delta_{q=\Delta/v_F}$.  This is the essence of the so called bifurcation
approximation. As a result, we arrive at the following equation:
\begin{equation}
\Delta_{p} -J \simeq \lambda
\left(\int_{\Delta/v_{F}}^{p} \frac{dq }{p} \Delta_{q}
+\int_{p}^{\Lambda} \frac{dq }{q} \Delta_{q} \right),
\end{equation}
where $\Lambda \simeq \pi/a$ is the ultraviolet cutoff ($a$ is a
lattice size). The last integral equation is equivalent to the
differential equation,
\begin{equation}
p^2\Delta_p^{\prime\prime}+2p\Delta_p^{\prime}+\lambda\Delta_p=0,
\label{diff_eq}
\end{equation}
with appropriate boundary conditions.
A non-trivial solution to the gap equation, 
satisfying the infrared boundary condition, 
\begin{equation}
\left. p^{2}\Delta_{p}^{\prime}\right|_{p=\Delta/v_{F}} =0,
\label{IRBC}
\end{equation}
exists only for $\lambda \geq 1/4$, and it takes the following form:
\begin{equation}
\Delta_{p} =\frac{\Delta^{3/2}}{\sin\delta \sqrt{v_{F} p}}
\sin\left(\frac{\nu}{2}\ln\frac{v_{F} p}{\Delta }+\delta\right),
\end{equation}
where $\nu = \sqrt{4\lambda-1}$ and $\delta = \arctan\nu$.
The critical value $\lambda_c=1/4$ corresponds to a continuous quantum
phase transition.

In addition, the ultraviolet boundary condition,
\begin{equation}
J=\left.(\Delta_{p}+ p\Delta_{p}^{\prime})\right|_{p=\Lambda},
\end{equation}
produces the relation:
\begin{equation}
J=\frac{\Delta^{3/2}}{\sin(2\delta) \sqrt{v_{F} \Lambda}}
\sin\left(\frac{\nu}{2}\ln\frac{v_{F} \Lambda}{\Delta}+2\delta
\right).
\label{J-del}
\end{equation}
This relation determines $J$ as a function of the gap $\Delta$. 
If one supplies it with 
a similar representation for the composite field $\sigma(\Delta)$, 
expression (\ref{eff-pot-def})
could be used directly to derive the effective potential,
\begin{equation}
V(\sigma(\Delta))=\int\limits^\Delta
d\Delta\frac{d\sigma(\Delta)}{d\Delta}J(\Delta).
\label{V-del}
\end{equation}
As follows from the definition, the expression for $\sigma$ as 
a function of $\Delta$ is given by
the trace of the quasiparticle propagator (multiplied by the 
factor $i$). Thus,
\begin{eqnarray}
\sigma &=& -\langle\bar\psi\psi\rangle=\frac{N_f}{\pi v_{F}}
\int_{0}^{\Lambda} \frac{q d q \Delta_q}
{\sqrt{q^{2}+(\Delta_q/v_{F})^{2}}}\nonumber\\
&=&-\left.
\frac{N_f}{\pi\lambda v_{F}}
p^2\Delta^\prime(p)\right|_{p=\Lambda} 
\nonumber\\
&=&\frac{N_f\Delta^{3/2}\sqrt{\Lambda}}
{\pi\lambda v_{F}^{3/2} \sin(2\delta)}
\sin\left(\frac{\nu}{2}\ln\frac{v_{F} \Lambda}{\Delta }
\right).
\label{sig-del}
\end{eqnarray}
Note that the second line of this equation follows from 
Eq.~(\ref{gap_eq:nonlin}) after differentiating its 
both sides with respect to momentum $p$ and 
substituting $p=\Lambda$.

In order to get the most convenient representation for the potential, 
we introduce a new parameter $\Delta_{0}$ which denotes the dynamical 
gap in the case of a vanishing external source. It is determined from
Eq.~(\ref{J-del}) at $J=0$:
\begin{equation}
\Delta_0=\Lambda v_{F}\exp\left[{-\frac{2\pi-4\delta}{\nu}}\right].
\end{equation}
Therefore the critical coupling $\lambda_c=1/4$ corresponds to a continuous 
phase transition of infinite order.

\begin{figure}[ht]
\centering{
\includegraphics[width=8cm]{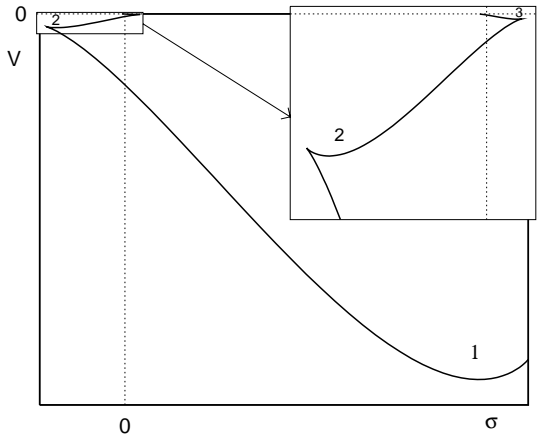}}
\caption{The schematic fractal structure of the effective potential 
$V(\sigma)$ that
appears as the consequence of the long range interaction.}
\label{fig-eff-fract}
\end{figure}

By making use of 
the last equation, we obtain a rather convenient representation 
of the functions $J(\Delta)$ and $\sigma(\Delta)$:
\begin{eqnarray}
J(\Delta)&=&-\frac{\Delta^{3/2}}{\sin(2\delta) \sqrt{v_{F} \Lambda}}
\sin\left(\frac{\nu}{2}\ln\frac{\Delta_{0}}{\Delta}\right),
\label{source-through-Delta} \\
\sigma(\Delta) &=& -\frac{N_f\Delta^{3/2}\sqrt{\Lambda}}
{\pi\lambda v_{F}^{3/2} \sin(2\delta)}
\sin\left(\frac{\nu}{2}\ln\frac{\Delta_{0}}{\Delta }-2\delta\right).
\label{sigma-through-Delta}
\end{eqnarray}
Finally, by substituting these into Eq.~(\ref{V-del}) and performing the 
integration on the right hand side, we arrive at the following parametric 
form of the effective potential:
\begin{eqnarray}
V(\Delta)& =& \frac{N_f\Delta^{3}}{2 \pi v_{F}^{2} } \Bigg\{
\frac{1-\nu^{2}}{2\nu^{2}}\left[1
-\cos\left(\nu\ln\frac{\Delta}{\Delta_{0}}\right)\right]
\nonumber\\
&+&\frac{1}{\nu}\sin\left(\nu\ln\frac{\Delta}{\Delta_{0}}\right)
-\frac{1}{3}\Bigg\},
\label{pot-through-Delta} \\
\sigma(\Delta)& =& \frac{4N_f\Delta^{3/2}\sqrt{\Lambda}}
{\pi (1+\nu^{2})v_{F}^{3/2} } \Bigg[
\cos\left(\frac{\nu}{2}\ln\frac{\Delta}{\Delta_{0}}\right) 
\nonumber\\
&+&\frac{1-\nu^{2}}{2\nu} 
\sin\left(\frac{\nu}{2}\ln\frac{\Delta}{\Delta_{0}}\right)
\Bigg].
\label{sig-through-Delta}
\end{eqnarray}
Around the global minimum, $\sigma_{0}=\sigma(\Delta_{0})$,
the potential is approximated as
\begin{equation}
V(\sigma) = \frac{\pi (1+\nu^{2})^{2} v_{F} \sigma^{2}}{48 N_{f} \Lambda}
\left(\ln\frac{\sigma}{\sigma_{0}}-\frac{1}{2}\right).
\end{equation}
This expression may suggest that there is nothing unusual 
about the potential given by the parametric representation in 
Eqs.~(\ref{pot-through-Delta}) and (\ref{sig-through-Delta}).
A closer look, however, reveals
a rather rich, fractal, structure of $V(\sigma)$ with infinite 
number of branches near the origin $\sigma=0$ 
(see Fig.~\ref{fig-eff-fract} and discussion below).

The extrema of the potential $V(\sigma)$ are determined
by the equation $dV/d \sigma = J(\Delta)=0$. By solving this
equation, we obtain an infinite set of solutions for $\Delta$:
\begin{equation}
\Delta_{\rm min}^{(n)} = \Delta_{0} 
\exp\left(-\frac{2\pi n}{\nu}\right), \quad n=0,1,\dots ,
\label{Delta_min_n}
\end{equation}
The corresponding values of $\sigma(\Delta_{\rm min}^{(n)})$ are 
determined by Eq.~(\ref{sig-through-Delta}). It is easy to check 
that the second derivative of the potential calculated at every 
extremal point in Eq.~(\ref{Delta_min_n}) is positive, 
${d^2V}/{d\sigma^2}>0$. Therefore, these extrema are local minima.
The values of the potential at these minima
$\sigma(\Delta_{\rm min}^{(n)})$ are calculated to be
\begin{equation}
V(\Delta_{\rm min}^{(n)})=-\frac{N_f}{6\pi v_F^2}
\left(\Delta_{\rm min}^{(n)}\right)^3<0.
\end{equation}
We see that the solution with $n=0$ gives the largest value of the
fermion gap and the lowest value of the potential. Therefore, it
corresponds to the global minimum of $V(\sigma)$, or in other words,
to the most stable (ground) state in the model at hand.

It is natural to expect that the potential also has maxima lying between 
those minima. However, the situation is more subtle: while, as a function
of parameter $\Delta$, the potential does have maxima, it does not
have them as a function of $\sigma$. Instead, the potential $V(\sigma)$
has turning points there, see Fig.~\ref{fig-eff-fract}. Indeed, 
the first derivative of the potential (\ref{pot-through-Delta}) 
with respect to the parameter 
$\Delta$ is zero at the following maxima: 
\begin{equation}
\Delta_{\rm trn}^{(n)} = \Delta_{0} \exp\left(-\frac{2\pi n}{\nu}
-\frac{2}{\nu}\arctan\frac{\nu (7-\nu^{2})}{3-5\nu^{2}}\right),
\label{Delta_trn_n}
\end{equation}
(with $n=0,1,\dots $). At the same time, the first derivative of 
$\sigma$-field with respect to $\Delta$ is also zero at these points.
As a result, the derivative of the potential with respect to $\sigma$,
i.e., $dV/d\sigma\equiv \frac{dV}{d\Delta}/\frac{d\sigma}{d\Delta}$, is
nonzero there.

From our analysis, we see that the potential $V(\sigma)$, defined
parametrically in Eqs.~(\ref{pot-through-Delta}) and
(\ref{sig-through-Delta}), is a multi-branched and multi-valued function
of $\sigma$. The location of the branching (turning) points are determined
by the values of the gap parameter given in Eq.~(\ref{Delta_trn_n}). Each
branch of the effective potential has a local minimum at $\sigma = \sigma
\left(\Delta_{\rm min}^{(n)}\right)$.  We also notice that, while the
locations of the branching points converge to $\sigma=0$ in the limit
$n\to\infty$, the shape of all higher branches (after an overall scaling
transformation)  resembles the shape of the first branch. Therefore, the
potential around the origin exhibits a {\em fractal} structure. This is
shown schematically in Fig.~\ref{fig-eff-fract}.

We will now demonstrate that the fractal structure of the potential
reflects a
rich spectrum of excitonic composites in this model. It is well known
(see, for example, Ref.~\cite{book}) that, because of the Ward identities
for currents connected with spontaneously broken symmetries, the gap
equation coincides with the Bethe-Salpeter equation for corresponding
gapless Nambu-Goldstone composites (gapless excitons in the present case).
Therefore, the infinite number of the solutions $\Delta_{\rm min}^{(n)}$
in Eq.~(\ref{Delta_min_n}) for the gap implies that there are gapless
excitons in each of the vacua corresponding to different values of $n$.
The genuine, stable, vacuum is of course that with $n=0$. Let us show that
the excitonic composites, which are gapless in the false vacua with $n >
0$, become gapped in the genuine vacuum with $n=0$. Indeed, the transition
from a false vacuum (with $n > 0$) to the genuine one corresponds to
increasing the fermion gap, $\Delta_{\rm min}^{(n)} \to \Delta_{\rm
min}^{(0)} \equiv \Delta_{0}$, {\it without} changing the dynamics.
Therefore, as a result of this increase of the gap of their constituents,
these excitonic composites should become gapped. Thus we conclude that the
fractal structure of the potential reflects the presence of an infinite
tower of gapped excitons that are radial excitations of the gapless
Nambu-Goldstone 
composites. It is clear that this effect is intimately connected with the
long-range interactions in this model.

A confirmation of this point comes from studying the effective potential
at finite temperature $T$ and/or chemical potential $\mu$. Because of the
Debye screening, both $T$ and $\mu$ lead to a dynamical infrared cutoff
for interactions. Our studies show that the (dis-)appearance of the
higher order branches is very sensitive to the values of temperature and
the chemical potential. In particular, all branches with $n>n_0$ disappear 
at a temperature in the region $\frac{1}{2}\Delta_{\rm min}^{(n_0+1)} \lesssim
T \lesssim \frac{1}{2}\Delta_{\rm min}^{(n_0)}$, or at a chemical potential
in the region $\sqrt{2}\Delta_{\rm min}^{(n_0+1)} < \mu \leq \sqrt{2}
\Delta_{\rm min}^{(n_0)}$.
This observation reflects the process of ``melting" of those
bound states whose binding energy is of order $T$ (or $\mu$) or less.

It is also clear that this picture should be quite general and valid for a
wide class of models with long-range interactions. In particular, we have
found that a similar fractal structure takes place in the effective action
of QED$_{2+1}$ relevant for cuprates.

The work of E.V.G. was supported by the research grant from FAPEMIG.
V.P.G. and V.A.M. are grateful for support from the Natural Sciences and
Engineering Research Council of Canada.
The research of V.P.G. has been also supported in part by NSF Grant No.
PHY-0070986 and by the SCOPES-projects 7~IP~062607 and 7UKPJ062150.00/1 of
Swiss NSF. The work of I.A.S. was supported by Gesellschaft f\"{u}r
Schwerionenforschung (GSI) and by Bundesministerium f\"{u}r Bildung und
Forschung (BMBF).

\end{document}